\documentclass[a4paper]{article}

\usepackage{INTERSPEECH2018}

\title{Attention-based End-to-End Models for Small-Footprint Keyword Spotting}
\name{Changhao Shan$^{1,2}$, Junbo Zhang$^2$, Yujun Wang$^2$, Lei Xie$^{1*}$}
\address{
  $^1$Shaanxi Provincial Key Laboratory of Speech and Image Information Processing, \\
  School of Computer Science, Northwestern Polytechnical University, Xi'an, China \\
  $^2$Xiaomi Inc., Beijing, China}
\email{{\{chshan, lxie\}}@nwpu-aslp.org, {\{zhangjunbo, wangyujun\}}@xioami.com}

\begin{document}

\maketitle
\begin{abstract}
	In this paper, we propose an attention-based end-to-end neural approach for small-footprint keyword spotting (KWS), which aims to simplify the pipelines of building a production-quality KWS system. Our model consists of an encoder and an attention mechanism. The encoder transforms the input signal into a high level representation using RNNs. Then the attention mechanism weights the encoder features and generates a fixed-length vector. Finally, by linear transformation and softmax function, the vector becomes a score used for keyword detection. We also evaluate the performance of different encoder architectures, including LSTM, GRU and CRNN. Experiments on real-world wake-up data show that our approach outperforms the recent Deep KWS approach by a large margin and the best performance is achieved by CRNN. To be more specific, with $\sim$84K parameters, our attention-based model achieves 1.02\% false rejection rate (FRR) at 1.0 false alarm (FA) per hour.
	
\end{abstract}

    
\section{Introduction}

    Keyword spotting (KWS), or spoken term detection (STD), is a task to detect pre-defined keywords in a stream of audio. Specifically, as a typical application of KWS, wake-up word detection has become an indispensable function on various devices, in order to enable users to have a fully hands-free experience. A practical on-device KWS module must minimize the false rejection rate at a low false alarm rate to make it easy to use, while limiting the memory footprint, latency and computational cost as small as possible.
    
    As a classic solution, large vocabulary continuous speech recognition (LVCSR) based systems~\cite{motlicek2012improving, can2011lattice} are widely used in the KWS task. Although it is flexible to change keywords according to user's requirement, the LVCSR based systems need to generate rich lattices and high computational resources are required for keyword search. These systems are often designed to search large databases of audio content. Several recent attempts have been proposed to reduce the computational cost, e.g., using end-to-end based acoustic models~\cite{bai2016end, Rosenberg2017End}. But these models are still quite large, making them not suitable for small-footprint, low-latency applications.
    Another classic technique for KWS is the keyword/filler hidden Markov model (HMM) approach~\cite{rohlicek1989continuous}, which remains strongly competitive until today. HMMs are trained for both keyword and non-keyword audio segments, respectively. At runtime, Viterbi decoding is used to search the best path in the decoding graph, which can be computationally expensive depending on the HMM topology. In these approaches, Gaussian mixture models (GMMs) were originally used to model the observed acoustic features, but with the advances in deep learning, deep neural networks (DNNs) have been recently adopted to substitute GMMs~\cite{Sz2005Comparison} with improved performances. Some studies replaced HMM by an RNN model trained with connectionist temporal classification (CTC) criterion~\cite{hwang2015online} or by an attention-based model~\cite{he2017streaming}, however, these studies are still under the keyword/filler framework.
    
    As a small footprint approach used by Google, Deep KWS~\cite{chen2014small} has drawn much attention recently. In this approach, a simple DNN is trained to predict the frame-level posteriors of sub keyword targets and fillers. When a confidence score, produced by a posterior handing method, exceeds a threshold, a keyword is detected. With no HMM involved, this approach has shown to outperform a keyword/filler HMM approach. In addition, this approach is highly attractive to run on the device with small footprint and low latency, as the size of the DNN can be easily controlled and no graph-searching is involved. Later, feed-forward DNNs were substituted by more powerful networks like convolutional neural networks (CNNs)~\cite{sainath2015convolutional} and recurrent neural networks (RNNs)~\cite{sun2016max}, with expected improvements. It should be noted that, although the framework of Deep KWS is quite simple, it still needs a well-trained acoustic model to obtain frame-level alignments.
    
    In this paper, we aim to further simplify the pipelines of building a production-quality KWS. Specifically, we propose an attention-based end-to-end neural model for small-footprint keyword spotting. By saying \textit{end-to-end}, we mean that: (1) a simple model that directly outputs keyword detection; (2) no complicated searching involved; (3) no alignments needed beforehand to train the model. Our work is inspired by the recent success of attention models used in speech recognition~\cite{bahdanau2016end, chan2016listen, shanattention}, machine translation~\cite{bahdanau2014neural}, text summarization~\cite{rush2015neural} and speaker verification~\cite{chowdhury2017attention}. It is intuitive to use attention mechanism in KWS: humans are able to focus on a certain region of an audio stream with ``high resolution" (e.g., the listener's name) while perceiving the surrounding audio in ``low resolution", and then adjusting the focal point over time. 

    Our end-to-end KWS model consists of an \textit{encoder} and an \textit{attention} mechanism. The encoder transforms the input signal into a high level representation using RNNs. Then the attention mechanism weights the encoder features and generates a fixed-length vector. Finally, by linear transformation and softmax function, the vector becomes a score used for keyword detection. In terms of end-to-end and small-footprint, the closest approach to ours is the one proposed by Kliegl \textit{et al.}~\cite{arik2017convolutional}, where a convolutional recurrent neural network (CRNN) architecture is used. However, the latency introduced by its long decoding window ($T$=1.5 secs) makes the system difficult to use in real applications. 

    To improve our end-to-end approach, we further explore the encoder architectures, including LSTM~\cite{hochreiter1997long}, GRU~\cite{cho2014learning} and CRNN that is inspired by~\cite{arik2017convolutional}. Experiments on real-world wake-up data show that our approach outperforms Deep KWS by a large margin. GRU is preferred over LSTM and the best performance is achieved by CRNN. To be more specific, with only $\sim$84K parameters, the CRNN-based attention model achieves 1.02\% false rejection rate (FRR) at 1.0 false alarm (FA) per hour.
    
\section{Attention-based KWS}

    \subsection{End-to-end architecture}
    
    \begin{figure}[t]
    	\centering
    	\includegraphics[width=\linewidth]{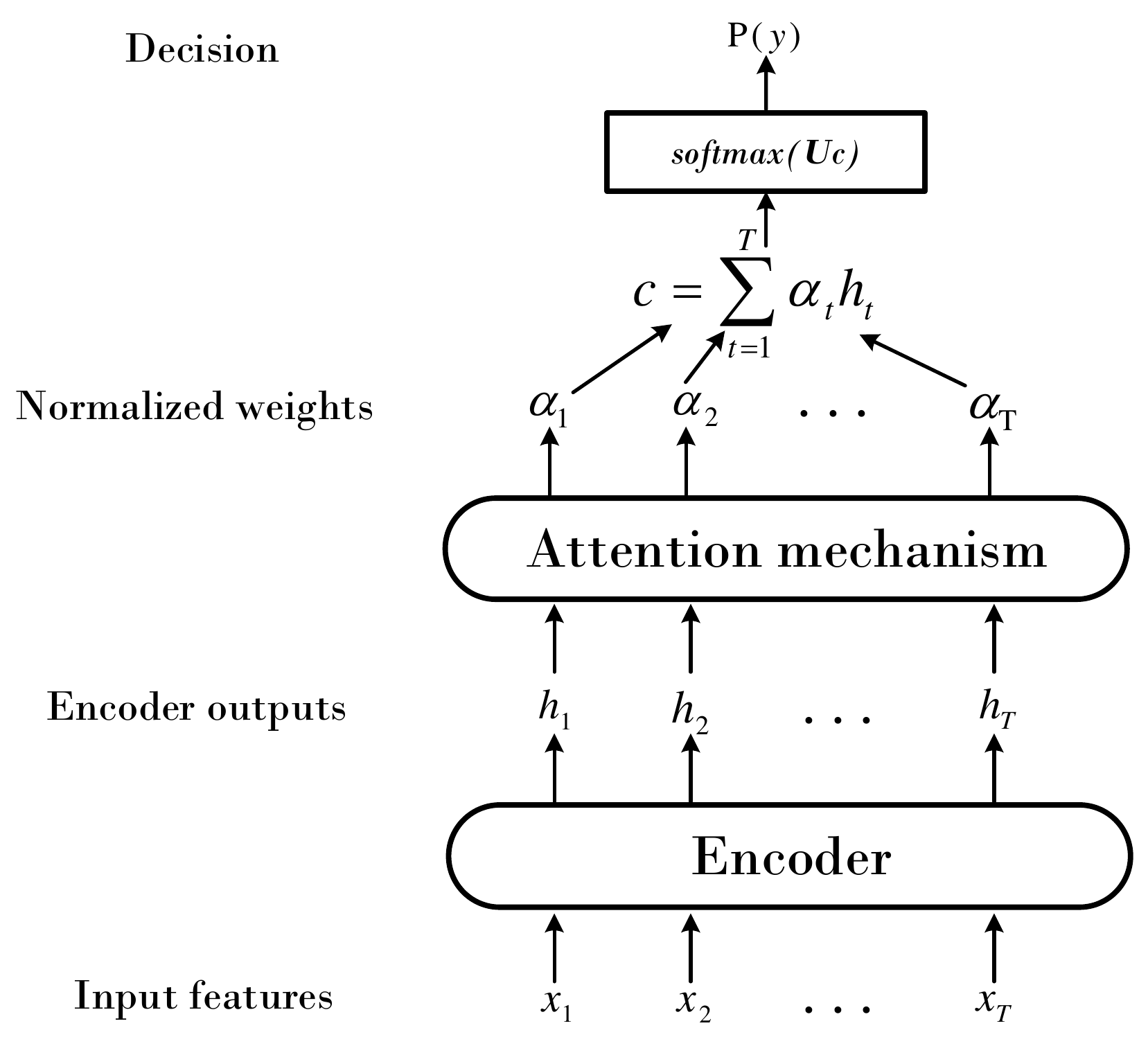}
    	\caption{Attention-based end-to-end model for KWS.}\vspace{-10pt}
    	\label{fig:end2end}
    \end{figure}
    
   We propose to use attention-based end-to-end model in small-footprint keyword spotting. As depicted in Fig.~\ref{fig:end2end}, the end-to-end architecture consists of two major sub-modules: the encoder and the attention mechanism. The encoder results in a higher-level feature representation $\mathbf{h} = (h_1,...,h_T)$ from the input speech features $\mathbf{x} = (x_1,...,x_T)$:
    \begin{flalign}
    	\mathbf{h}&=Encoder(\mathbf{x}).
    	\label{eq1}
    \end{flalign}
    Specifically, the $Encoder$ is usually a RNN that can directly make use of speech contextual information. In our work, we explore different encoder structures, including GRU, LSTM and CRNN. The attention mechanism learns normalized weights $\alpha_t \in [0,1]$ from the feature representation:
    \begin{flalign}
        \alpha_{t}&=Attend(\bm{h}_t).
        \label{eq2}
    \end{flalign}
    Then we form fixed-length vector $c$ as the weighted average of the $Encoder$ outputs $\mathbf{h}$:
    \begin{flalign}
    \bm{c}&=\sum_{t=1}^T\alpha_{t}\bm{h}_t.
    \label{eq3}
    \end{flalign}
    Finally, we generate a probability distribution by a linear transformation and the softmax function:
    \begin{flalign}
    p(y)&=softmax(\bm{U}\bm{c}).
    \label{eq4}
    \end{flalign}
    where $\mathbf{U}$ is the linear transform, $y$ indicate whether a keyword detected.

    \subsection{Attention mechanism}
    
    Similar to human listening attention, the attention mechanism in our model selects the speech parts which are more likely to contain the keyword while ignoring the unrelated parts. We investigate both average attention and soft attention.
    
    \textbf{Average attention}: The $Attend$ model does not have trainable parameters and the $\alpha_t$ is set as the average of $T$:
    \begin{flalign}
    \alpha_{t}&=\frac{1}{T}.
    \label{eq5}
    \end{flalign}
     
     \textbf{Soft attention}: This attention method is borrowed from speaker verification~\cite{chowdhury2017attention}. Compared with other attention layers, the shared-parameter non-linear attention is proven to be effective~\cite{chowdhury2017attention}. We first learn a scalar score $e_t$:
     \begin{flalign}
     e_{t}&=v^Ttanh(\bm{W}h_t+\bm{b}).
     \label{eq6}
     \end{flalign}
     Then we compute the normalized weight $\alpha_t$ using these scalar scores:
     \begin{flalign}
     \alpha_t&=\frac{exp(e_t)}{\sum_{j=1}^Texp(e_j)}.
     \label{eq7}
     \end{flalign}
    
    \subsection{Decoding}
    
    \begin{figure}[t]
    	\centering
     	\includegraphics[width=\linewidth]{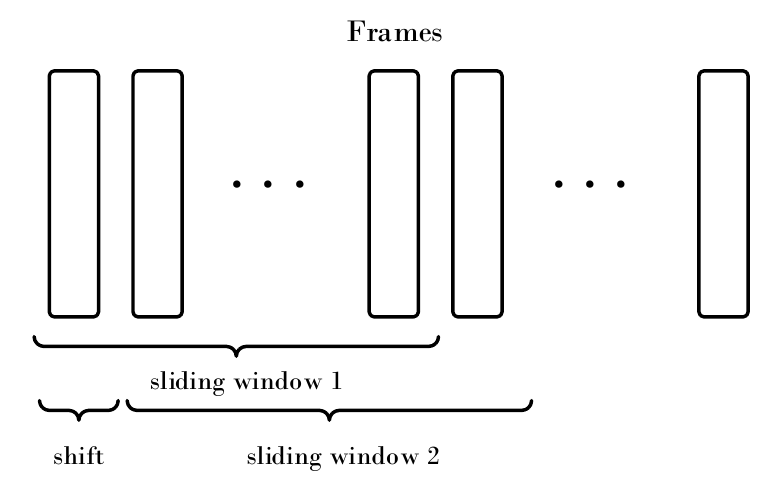}
    	\caption{Sliding windows used in decoding.}\vspace{-10pt}
    	\label{fig:decoding}
    \end{figure}
    
    As shown in Fig.~\ref{fig:end2end}, unlike some other approaches~\cite{chen2014small}, our end-to-end system outputs a confidence score directly without post-processing. Similar to the Deep KWS system, our system is triggered when the $p(y=1)$ exceeds a preset threshold. During decoding, in Fig.~\ref{fig:decoding}, the input is a sliding window of speech features, which has a preset length and contains the entire keyword. Meanwhile, a frame shift is employed. The small set of parameters in our system leads to small-footprint memory use. For a sliding window, we only need to feed one frame into the network for computation and the rest frames have been already computed in the previous sliding window. Therefore, our system has a low computational cost.
    
\section{Experiments}

    \begin{table}[t]
    	\caption{\label{tab:attention} {\it Performance comparison between Deep KWS and attention-based models with 2-64 network. FRR is at 1.0 false alarm (FA) per hour.}}
    	\vspace{2mm}
    	\footnotesize
    	\centerline{
    		\begin{tabular}{ l c  c }      			
    			\hline
    			\textbf{Model} & \textbf{FRR (\%)} & \textbf{Params (K)} \\
    			\hline \hline 
    			DNN KWS                & $13.9$          &   $62.5$ \\
    			\hline
    			LSTM KWS               & $7.10$          &   $54.1$ \\ 
    			LSTM average attention    & $4.43$          &   $60.0$ \\
    			LSTM soft attention     & $3.58$          &   $64.3$ \\
    			\hline
    			GRU KWS                & $6.38$          &   $44.8$ \\
    			GRU average attention     & $3.22$          &   $49.2$ \\
    			GRU soft attention     & $\textbf{1.93}$ &   $53.4$ \\
    			\hline
    		\end{tabular}
    	}\vspace{-10pt}
    \end{table}
    
    \begin{figure}[t]
    	\centering
    	\includegraphics[width=\linewidth]{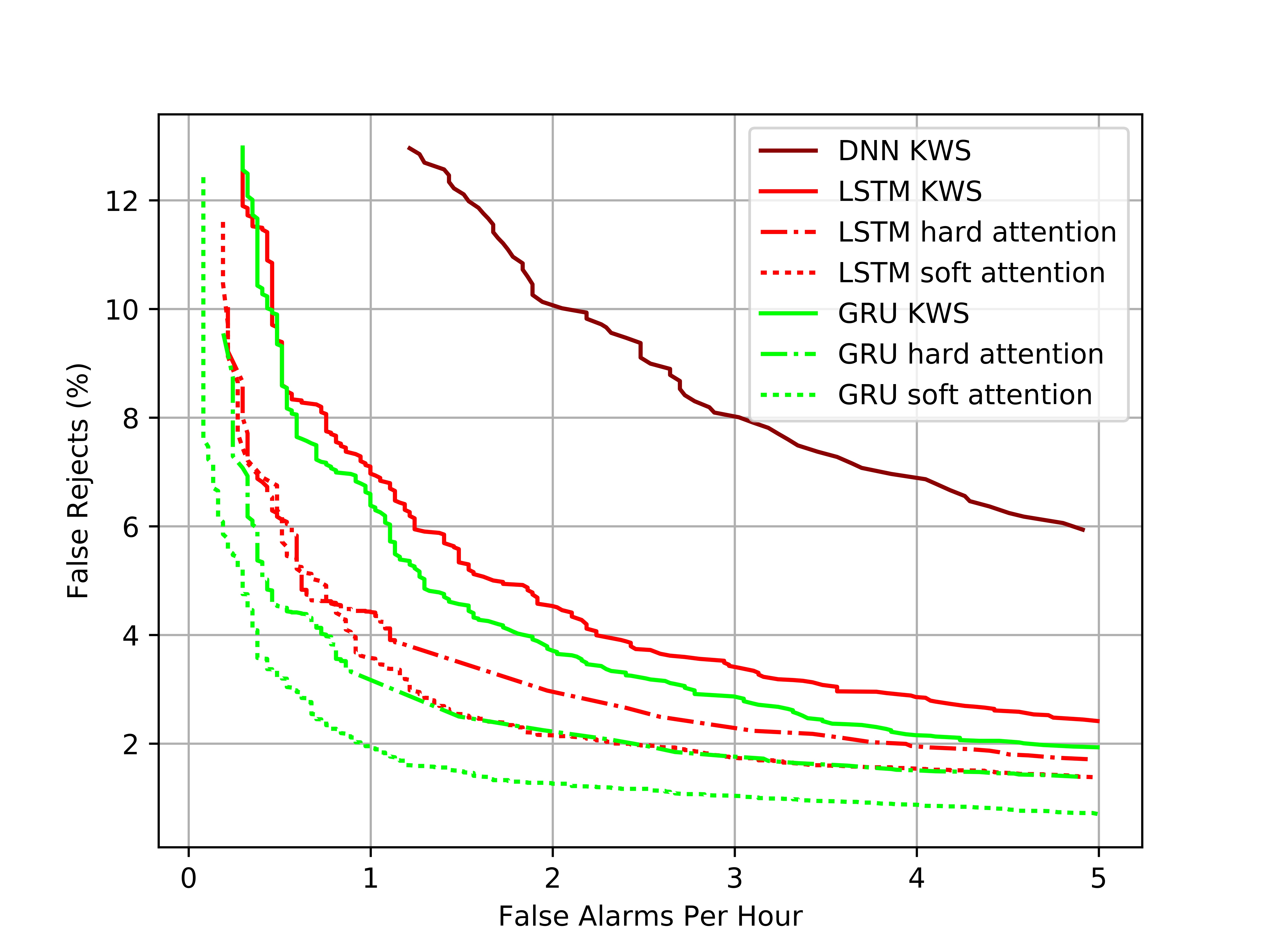}
    	\caption{ROCs for Deep KWS vs. Attention-based system with 2-64 network.}\vspace{-10pt}
    	\label{fig:attention}
    \end{figure}
    
    \subsection{Datasets}

    We evaluated the proposed approach using real-world wake-up data collected from Mi AI Speaker\footnote{https://www.mi.com/aispeaker/}. The wake-up word is a four-syllable Mandarin Chinese term (``xiao-ai-tong-xue"). We collected $\sim$188.9K positive examples ($\sim$99.8h) and $\sim$1007.4K negative examples ($\sim$1581.8h) as the training set. The held-out validation set has $\sim$9.9K positive examples and $\sim$53.0K negative examples. The test data set has $\sim$28.8K positive examples ($\sim$15.2h) and $\sim$32.8K negative examples ($\sim$37h). Each audio frame was computed based on a 40-channel Mel-filterbank with 25ms windowing and 10ms frame shift. Then the filterbank feature was converted to per-channel energy normalized (PCEN)~\cite{wang2017trainable} Mel-spectrograms.
    
    \subsection{Baseline}
    
    We reimplemented the Deep KWS system~\cite{chen2014small} as the baseline, in which the network predicts the posteriors for the four Chinese syllables in the wake-up word and a filler. The ``filler'' here means any voice that is not contain the keyword. Specifically, we adopted three different networks, including DNN, LSTM and GRU. For a fare comparison, the network configuration was set to have similar size of parameters with the proposed attention models. The feed-forward DNN model had 3 hidden layers and 64 hidden nodes per layer with rectified linear unit (ReLU) non-linearity. An input window with 15 left frames and 5 right frames was used. The LSTM and GRU models were built with 2 hidden layers and 64 hidden nodes per layer. For the GRU KWS model, the final GRU layer was followed by a fully connected layer with ReLU non-linearity. There were no stacked frames in the input for the LSTM and GRU models. The smoothing window for Deep KWS was set to 20 frames. We also trained a TDNN-based acoustic model using $\sim$3000 hours of speech data to perform frame-level alignment before KWS model training.    
    
    \subsection{Experimental Setup}
    
    In the neural network models, all the weight matrices were initialized with the normalized initialization~\cite{Glorot2010Understanding} and the bias vectors were initialized to 0. We used ADAM~\cite{kingma2014adam} as the optimization method while we decayed the learning rate from 1e-3 to 1e-4 after it converged. Gradient norm clipping to 1 was applied, together with L2 weight decay 1e-5. The positive training sample has a frame length of $T=1.9$ seconds which ensures the entire wake-up word is included. Accordingly, in the attention models, the input window has set to 189 frames to cover the length of the wake-up word. We randomly selected 189 contiguous frames from the negative example set to train the attention models. At runtime, the sliding window was set to 100 frames and frame shift was set to 1. Performances were measured by observing the FRR at the operating threshold of 1.0 FA per hour, while plotting a receiver operating curve (ROC).
    
    \subsection{Impact of attention mechanism}
    
    From Table~\ref{tab:attention} and Fig.~\ref{fig:attention}, we can clearly see the superior performances of the attention models. With similar size of parameters, the proposed attention models outperform the Deep KWS systems by a large margin. We also note that GRU is preferred over LSTM in both Deep KWS and the attention models. Not surprisingly, the soft attention-based model achieves the best performance. At 1.0 FA/hour, the GRU attention model reduces the FRR from 6.38\% (GRU Deep KWS) down to 1.93\% with a remarkable false rejection reduction.
    
    \begin{table}[t]
    	\caption{\label{tab:RNN} {\it Performance of different encoder architectures with soft attention. FRR is at 1.0 false alarm (FA) per hour.}}
    	\vspace{2mm}
    	\footnotesize
    	\centerline{
    		\begin{tabular}{ c c c c c }      			
    			\hline
    			\textbf{Recurrent Unit} & \textbf{Layer} & \textbf{Node} & \textbf{FRR (\%)} & \textbf{Params (K)} \\
    			\hline \hline
    			LSTM & $1$ & $64 $ & $4.36  $        &   $31.2$ \\
    			LSTM & $2$ & $64 $ & $3.58  $        &   $64.3$ \\ 
    			LSTM & $3$ & $64 $ & $3.05  $        &   $97.3$ \\
    			LSTM & $1$ & $128$ & $2.99  $        &   $103 $  \\
    			\hline
    			GRU  & $1$ & $64 $ & $3.22  $        &   $28.7$ \\
    			GRU  & $2$ & $64 $ & $1.93  $        &   $53.4$ \\
    			GRU  & $3$ & $64 $ & $1.99  $        &   $78.2$ \\
    			GRU  & $1$ & $128$ & $\textbf{1.49}$ &   $77.5$ \\
    			\hline
    		\end{tabular}
    	}\vspace{-10pt}
    \end{table}
    
    \begin{table}[t]
    	\caption{\label{tab:CRNN} {\it Performance of adding convolutional layers in the GRU (CRNN) attention-based model with soft attention. FRR is at 1.0 false alarm (FA) per hour.}}
    	\vspace{2mm}
    	\footnotesize
    	\centerline{
    		\begin{tabular}{ c c c c c c }      			
    			\hline
    			\textbf{Channel} & \textbf{Layer} & \textbf{Node} & \textbf{FRR (\%)} & \textbf{Params (K)} \\
    			\hline \hline
    			$8 $ & $1$ & $64$ & $2.48$          &   $52.5$ \\
    			$8 $ & $2$ & $64$ & $1.34$          &   $77.3$ \\
    			$16$ & $1$ & $64$ & $\textbf{1.02}$ &   $84.1$ \\
    			$16$ & $2$ & $64$ & $1.29$          &   $109 $ \\
    			\hline
    		\end{tabular}
    	}\vspace{-10pt}
    \end{table}
    
    \subsection{Impact of encoder architecture}
    
\begin{figure}[t]
   	\centering
    \includegraphics[width=\linewidth]{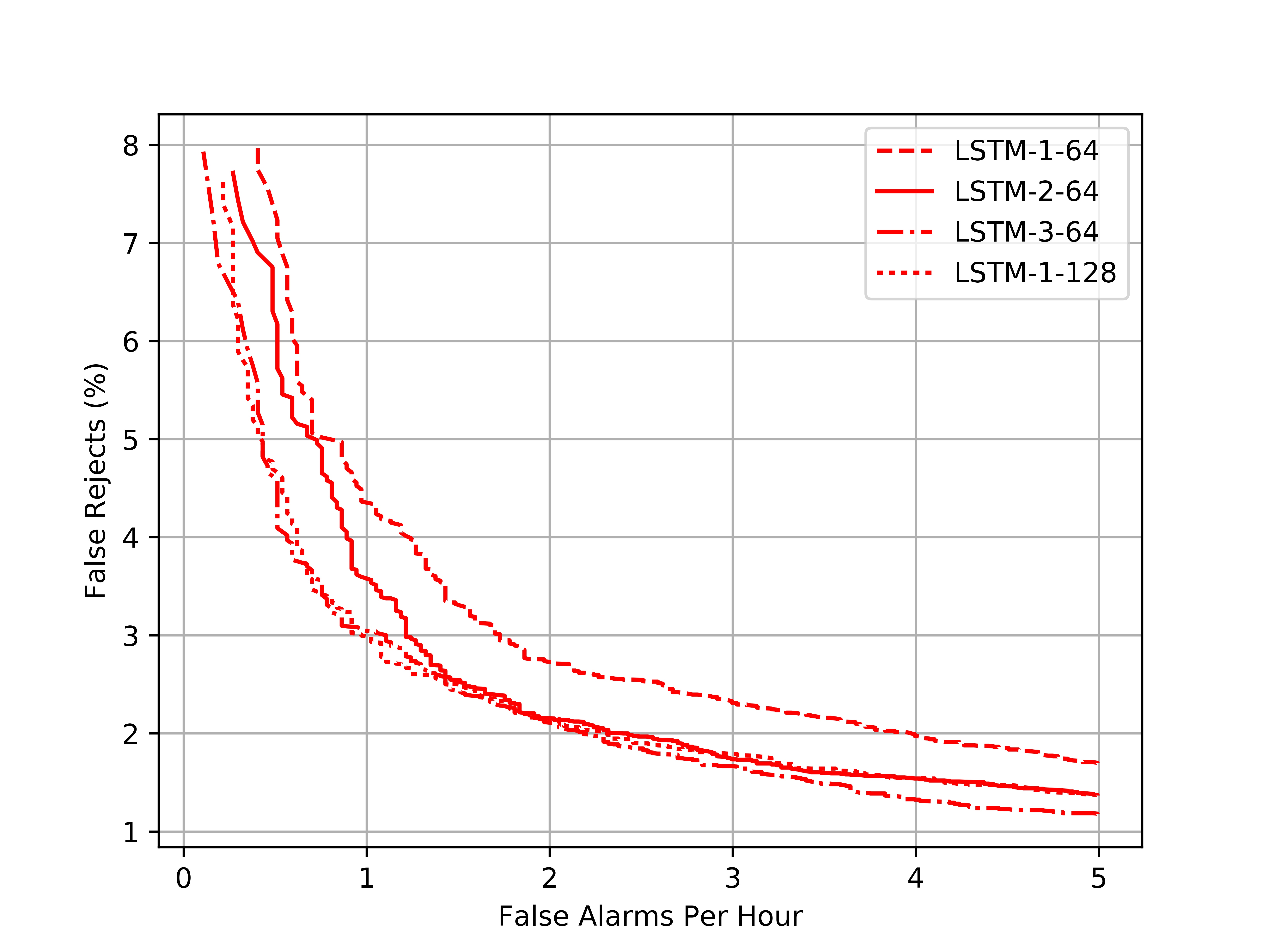}
    \caption{ROCs for LSTM Attention-based model with soft attention.}\vspace{-10pt}
    \label{fig:lstm-attention}
   \end{figure}

\begin{figure}[t]
    \centering
    \includegraphics[width=\linewidth]{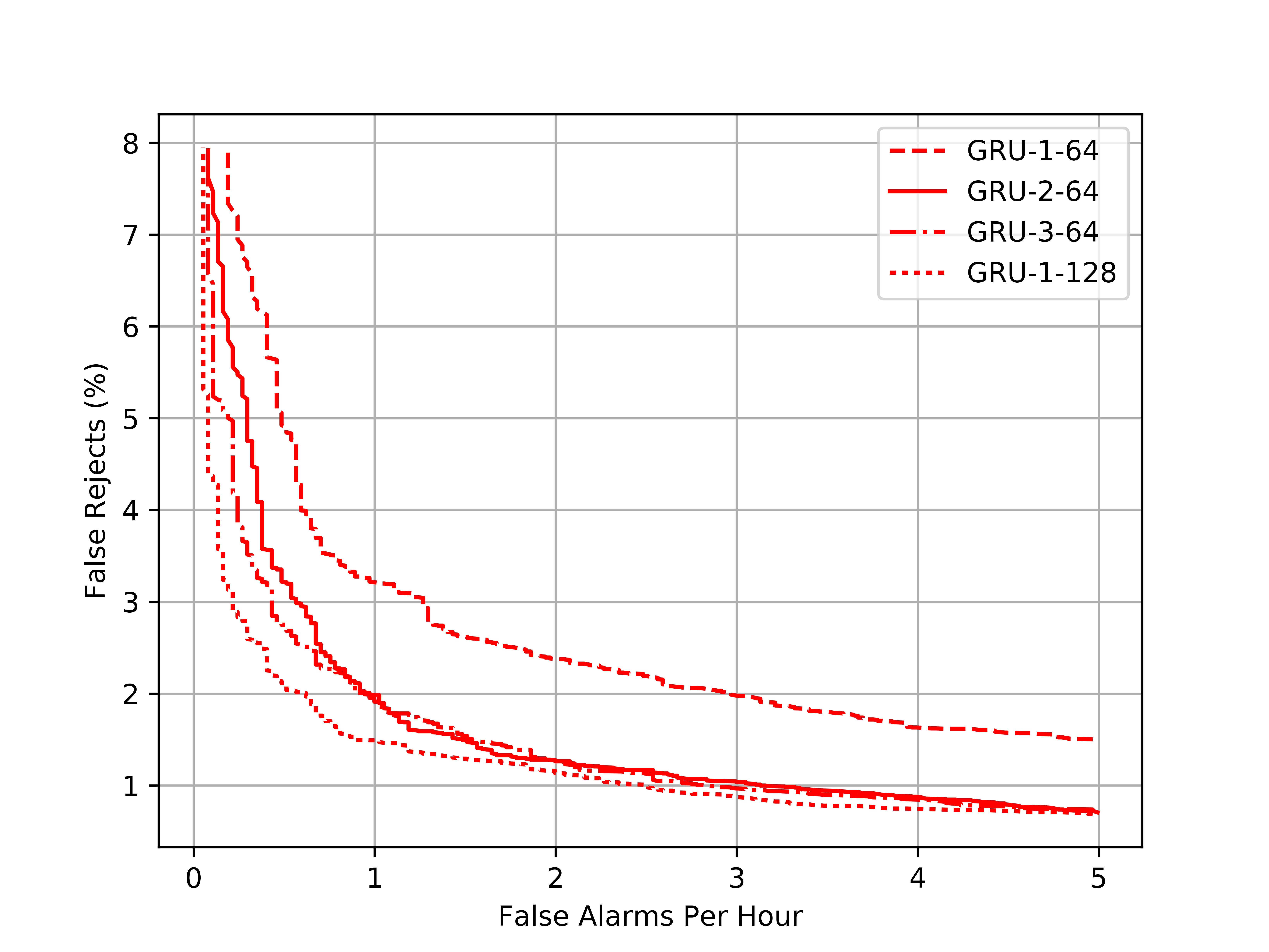}
    \caption{ROCs for GRU Attention-based model with soft attention.}\vspace{-10pt}
   	\label{fig:gru-attention}
   \end{figure}

    We further explored the impact of encoder architectures with soft attention. Results are summarized in Table~\ref{tab:RNN}, Fig.~\ref{fig:lstm-attention} and Fig.~\ref{fig:gru-attention}. From Table~\ref{tab:RNN}, we notice that the bigger models always perform better than the smaller models. Observing the LSTM models, the 1-128 LSTM model achieves the best performance with an FRR of 2.99\% at 1.0 FA/hour. In Fig.~\ref{fig:lstm-attention}, the ROC curves of the 1-128 LSTM model and the 3-64 LSTM model are overlapped at lower FA per hour. This means making the LSTM network wider or deeper can achieve the same effect.  However, observing Fig.~\ref{fig:gru-attention}, the same conclusion does not hold for GRU. The 1-128 GRU model presents a significant advantage over the 3-64 GRU model. In other words, increasing the number of nodes may be more effective than increasing the number of layers. Finally,  the 1-128 GRU model achieves 1.49\% FRR at 1.0 FA/hour.
 
  \subsection{Adding convolutional layer}
     
\begin{figure}[t]
    \centering
    \includegraphics[width=\linewidth]{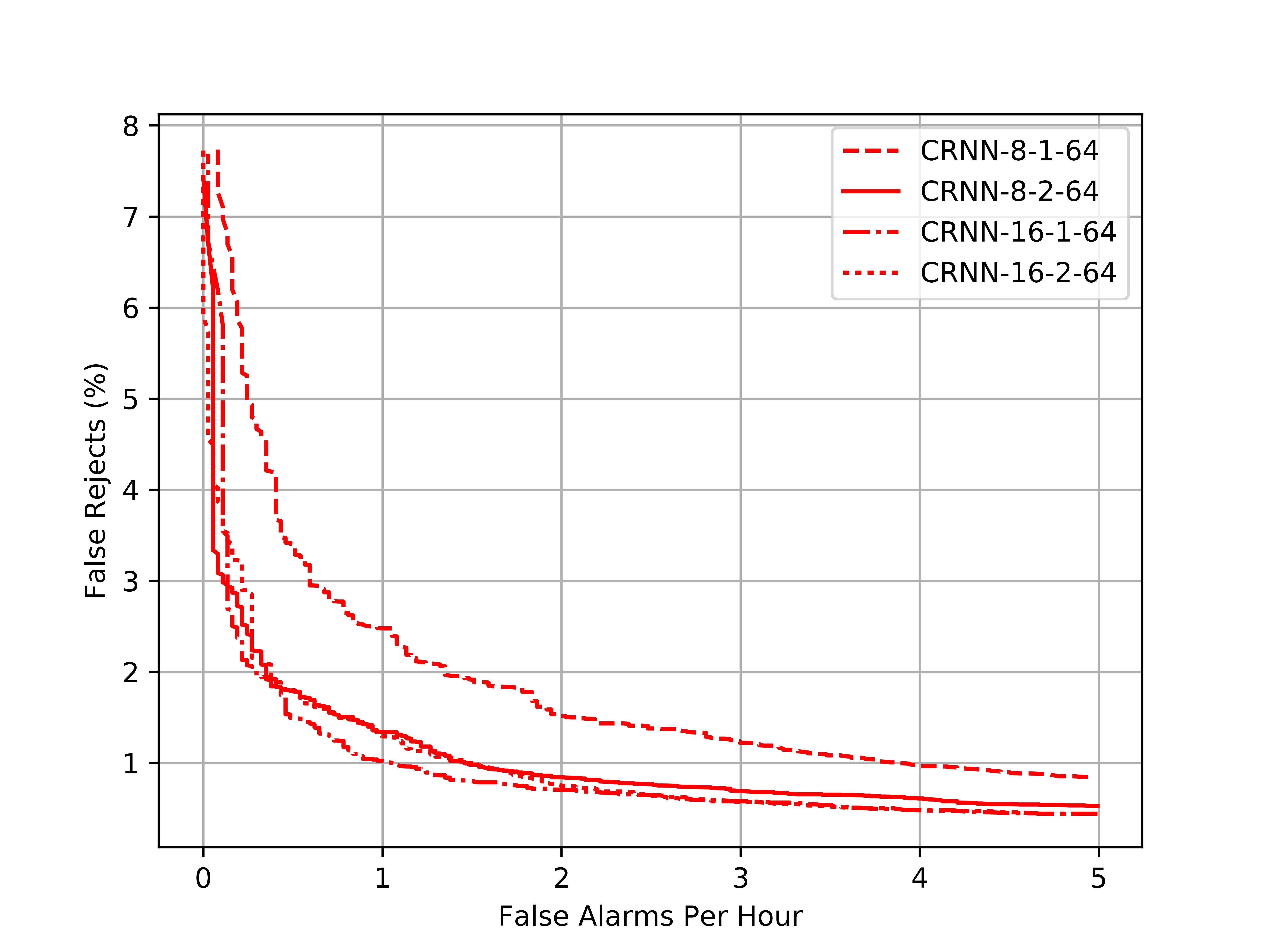}
    \caption{ROCs for CRNN Attention-based model with soft attention.}\vspace{-10pt}
    \label{fig:crnn-attention}
   \end{figure}

   Inspired by \cite{arik2017convolutional}, finally we studied the impact of adding convolutional layers in the GRU attention-model as convolutional network is often used as a way to extract invariant features. For the CRNN attention-based model, we used one layer CNN that has a $\frac{C(20\times5)}{1\times2}$ filter. We explored different numbers of output channel and results are summarized in Table~\ref{tab:CRNN} and Fig.~\ref{fig:crnn-attention}.  From Table~\ref{tab:CRNN}, we can see that adding convolutional layer can further improve the performance. We achieve the lowest FRR of 1.02\% at 1.0 FA/hour with 84.1K parameters. Another observation is that 16-channel models work better than 8-channel models. By increasing layers, the 8-2-64 model achieves a great gain over the 8-1-64 model. But we cannot observe extra benefit when increasing the layers with 16-channel models.
    
\begin{figure}[t]
    \centering
    \includegraphics[width=\linewidth]{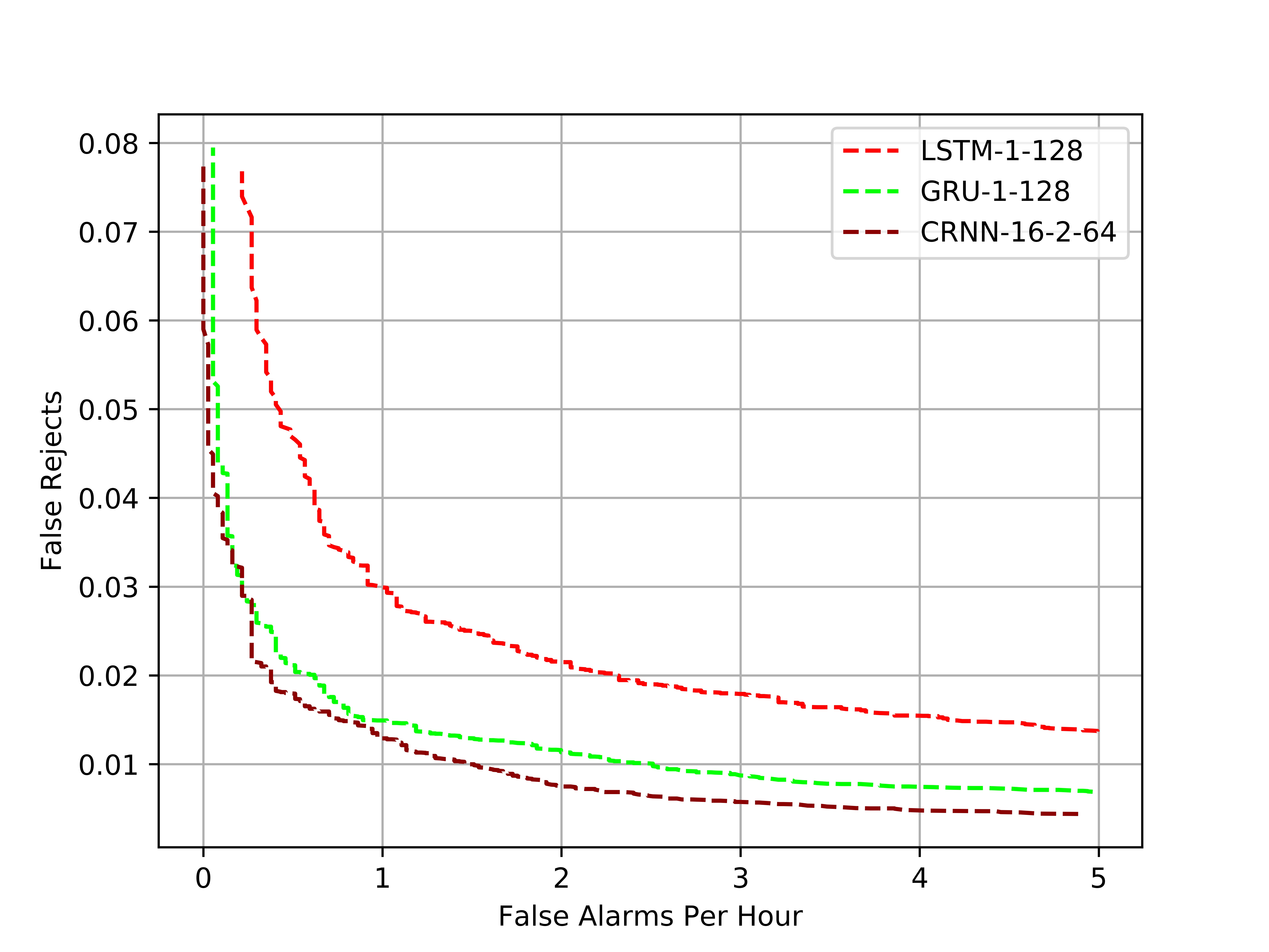}
    \caption{ROCs for different architectures with soft Attention-based model.}\vspace{-10pt}
    \label{fig:rnn-attention}
   \end{figure}

   As a summary, Fig.~\ref{fig:rnn-attention} plots the ROC curves for the best three systems. We can see that GRU and CRNN outperform LSTM by a large margin and the best performance is achieved by CRNN.
    
    
\section{Conclusions}

    In this paper, we propose an attention-based end-to-end model for small-footprint keyword spotting. Compared with the Deep KWS system, the attention-based system achieves superior performance. Our system consists of two main sub-modules: the encoder and the attention mechanism. We explore the encoder architectures, including LSTM, GRU and CRNN. Experiments show that GRU is preferred over LSTM and the best performance is achieved by CRNN. We also explore two attention mechanisms: average attention and soft attention. Our results show that the soft attention has a better performance than the average attention. With $\sim$84K parameters, our end-to-end system finally achieves 1.02\% FRR at 1.0 FA/hour. 

\section{Acknowledgements} 
    The authors would like to thank Jingyong Hou for helpful comments and suggestions.
\clearpage
\bibliographystyle{IEEEtran}

\bibliography{sch2018interspeech}

\end{document}